# Regularization Parameter Selection Method for Sign LMS with Reweighted L1-Norm Constriant Algorithm

Guan Gui and Li Xu

Department of Electronics and Information Systems, Akita Prefectural University, Yurihonjo, Akita 015-0055, Japan

E-mails: {guiguan,xuli}@akita-pu.ac.jp

## Abstract

Broadband frequency-selective fading channels usually have the sparsity nature. By exploiting the sparsity, adaptive sparse channel estimation (ASCE) algorithms, e.g., least mean square with reweighted L1-norm constraint (LMS-RL1) algorithm, could bring a considerable performance gain under assumption of additive white Gaussian noise (AWGN). In practical scenario of wireless systems, however, channel estimation performance is often deteriorated by unexpected non-Gaussian mixture noises which include AWGN and impulsive noises. To design stable communication systems, sign LMS-RL1 (SLMS-RL1) algorithm is proposed to remove the impulsive noise and to exploit channel sparsity simultaneously. It is well known that regularization parameter (REPA) selection of SLMS-RL1 is a very challenging issue. In the worst case, inappropriate REPA may even result in unexpected instable convergence of SLMS-RL1 algorithm. In this paper, Monte Carlo based selection method is proposed to select suitable REPA so that SLMS-RL1 can achieve two goals: stable convergence as well as usage sparsity information. Simulation results are provided to corroborate our studies.

**Keywords**: SLMS-RL1 algorithm; regularization parameter selection; adaptive sparse channel estimation; Gaussian mixture model (GMM).

1. **Introduction**

Broadband transmission is becoming more and more important in advanced wireless communications systems [1]–[3]. The main impairments in wireless systems are due to multipath propagation as well as harmful additive noises. In such circumstances, accurate channel state information (CSI) is required for stable coherence signal detection [4]. Based on the assumption of Gaussian noise model, second-order statistics based least mean square (LMS) algorithm and its variants have been widely developed to estimate channels due to its simplicity and robustness [5][6]. However, the performance of LMS is usually limited by potential impulsive noises in advanced wireless systems [7][8]. These kinds of impulsive noises are often generated from natural or man-made electromagnetic waves, usually has a long tail distribution and violates the commonly used Gaussian noise assumption [9]. Without loss of generality, Gaussian mixture noise model (GMM) has been used to describe non-Gaussian noise system [8].

To mitigate the harmful GMM noises, it is necessary to develop robust channel estimation algorithms. Based on the assumption of dense finite impulse response (FIR), recently, several effective adaptive channel estimation algorithms have been proposed to achieve the robustness against impulsive interferences [6][10]–[12]. In [6], standard sign least mean absolute (SLMS) is proposed to suppress impulsive noise with using sign LMS algorithm. In [10], an useful standard affine projection sign algorithm (APSA) is proposed to mitigate impulsive noise. In [11], Yoo et. al. propose an improved APSA algorithm by deriving approximate optimal step-size. In [12], Li et. al. propose an effective variable step-size (VSS) sign algorithm for stable channel estimation under Gaussian mixture noise environment. The performance gain is obtained by adjusting the step-size via gradient-based weighted average of the sign algorithm. However, FIR of the real wireless channel is often modeled as sparse or cluster-sparse and hence many of channel coefficients are zero [13]–[17]. Hence, these algorithms may not exploit the sparse structure information. Indeed, some potential performance gain could be obtained if adopting advanced adaptive channel estimation algorithms.

To exploit channel sparsity as well as to remove GMM noises, sign least mean square with reweighted L1-norm constraint (SLMS-RL1) [18] algorithm is proposed. It is well known that regularization parameter (REPA) is one of critical parameters to control the performance of SLMS-RL1 algorithm. This paper proposed a Monte Carlo based selection method to choose an appropriate REPA so that SLMS-RL1 algorithm can achieve estimation performance gain as much as possible and also can ensure stable convergence under different GMM noise levels. Simulations results are given to verify the effectiveness of the proposed algorithm.

The rest of the paper is organized as follows. In Section II, we introduce GMM-induced adaptive sparse system model and review SLMS-RL1 algorithm. In Section III, Monte Carlo based computer simulations are given to select REPA. Then, SLMS-RL1 algorithm using the proposed REPA is compared with benchmarking algorithms, i.e.

LMS, SLMS and LMS-RL1. Finally, Section IV concludes the paper and brings forward the future work.

## 2. Reviwe of RL1-SLMS Algorithm and Problem Formulation

Consider an additive noise interference channel, which is modeled by the unknown $N$-length finite impulse response (FIR) vector $\boldsymbol{w} = [w_0, w_1, \cdots, w_{N-1}]^T$ at discrete time index $n$. The ideal received signal is expressed as

$$d(n) = \boldsymbol{x}^T(n)\boldsymbol{w} + z(n), \tag{1}$$

where $\boldsymbol{x}(n) = [x(n), x(n-1), \cdots, x(n-N+1)]^T$ is the input signal vector of the $N$ most recent input samples; $\boldsymbol{w}$ is an $N$-dimensional column vector of the unknown system that we wish to estimate, and $z(n)$ is impulsive noise which can be described by Gaussian mixture model (GMM) [8] distribution as

$$p(z(n)) = (1-\phi) \cdot \mathcal{CN}(0, \sigma_n^2) + \phi \cdot \mathcal{CN}(0, T\sigma_n^2), \tag{2}$$

Where $T \gg 1$ denotes impulsive noise strength and $\mathcal{CN}(0, \sigma_n^2)$ denotes the Gaussian distributions with zero mean and variance $\sigma_n^2$, and $\phi$ is the mixture parameter to decide the level of impulsive noise. According to (2), one can find that stronger impulsive noises could be described by larger $T$ as well as larger $\phi$. Hence, variance of GMM noise $z(n)$ is obtained as

$$\sigma_z^2 = E(z^2(n)) = (1-\phi)\sigma_v^2 + \phi T \sigma_v^2. \tag{3}$$

Note that $z(n)$ reduces to Gaussian noise if $\phi = 0$. The objective of the adaptive channel estimation is to perform adaptive estimate of $w(n)$ with limited complexity and memory given sequential observation $\{d(n), x(n)\}$ in the presence of additive GMM noise $z(n)$. According to (1), instantaneous estimation error $e(n)$ can be written as

$$e(n) = d(n) - w^T(n)x(n), \tag{4}$$

where $w(n)$ is the estimator of $w$ at iteration $n$ and $v(n) = w(n) - w$. To estimate $w(n)$, RL1-SLMS algorithm is adopted. Firstly, the cost function is written as

$$G(n) = \|e(n)\|_1 + \lambda \|f(n)w(n)\|_1, \tag{5}$$

where $\lambda$ denotes a positive regularization parameter to balance estimation error and sparsity exploitation. In addition, the penalty term and elements of the $1 \times N$ row vector $f(n)$ are set to

$$[f(n)]_i = \frac{1}{\delta_r + |[w(n-1)]_i|}, \quad i = 0, 1, \cdots, N-1, \tag{6}$$

where $\delta_r$ being some positive number and hence $[f(n)]_i > 0$ for $i = 0, 1, ..., N-1$. The update equation can be derived by differentiating (5) with respect to the FIR channel vector $w(n)$. Then, the resulting update equation is:

$$\begin{aligned} w(n+1) &= w(n) + \mu \frac{\partial G(n)}{\partial w(n)} \\ &= w(n) + \mu x(n) \operatorname{sgn}(e(n)) - \frac{\rho \operatorname{sgn}(w(n))}{\delta_r + |w(n-1)|}, \end{aligned} \tag{7}$$

where $\rho = \mu\lambda$. In Eq. (7), since $\text{sgn}(f(n)) = \mathbf{1}_{1\times N}$, hence one can get $\text{sgn}(f(n)w(n)) = \text{sgn}(w(n))$. Note that although the weight vector $w(n)$ changes in every stage of this sparsity-aware SLMS-RL1 algorithm, it does not depend on $w(n)$, and the cost function $G(n)$ is convex. In (7), suitable REPA selection could exploit considerable sparity so that SLMS-RL1 obtains MSE performance gain. Inversely, insatiable REPA may cause instable convergence for SLMS-RL1. Hence, it is necessary to choose appropriate REPA.

### 3. Monte-Carlo bsaed REPA Selection Method and Numerical Simulations

In this section, the proposed SLMS-RL1 algorithm is evaluated in different scenarios: SNR, impulsive-noise strength $T$, mixture parameters $\phi$ as well as channel sparsity $K$. For achieving average performance, $M = 1000$ independent Monte-Carlo runs are adopted. The simulation setup is configured according to the typical broadband wireless communication system [3]. The signal bandwidth is 50MHz located at the central radio frequency of 2.1GHz. The maximum delay spread of $0.8\mu s$. Hence, the maximum length of channel vector $w$ is $N = 80$ and its number of dominant taps is set to $K \in \{2,4,8,16\}$. To validate the effectiveness of the proposed algorithms, average mean square error (MSE) standard is adopted. Channel estimators are evaluated by average MSE which is defined by

$$\text{MSE}\{w(n)\} = 10\log_{10}\left\{\frac{1}{M}\sum_{m=1}^{M}\left[\|w_m(n) - w_m\|_2^2 \big/ \|w_m\|_2^2\right]\right\} \tag{8}$$

where $w$ and $w(n)$ are the actual signal vector and reconstruction vector, respectively. The results are averaged over $M = 1000$ independent Monte-Carlo runs. Each dominant channel tap follows random Gaussian distribution as $\mathcal{CN}(0, \sigma_w^2)$ which is subject to $E\{\|w\|_2^2\} = 1$ and their positions are randomly decided within the $w$. The received SNR is defined as $P_0/\sigma_z^2$, where $P_0$ is the received power of the pseudo-random (PN) binary sequence for training signal. In addition, threshold parameter of SLMS-RL1 is set as $\delta_r = 0.05$ [19]. Detailed parameters for computer simulation are listed in Tab. 1.

TAB. 1. SIMULATION PARAMETERS.

| Parameters | Values |
|---|---|
| Training signal | Pseudo-random Binary sequences |
| Channel length | $N = 80$ |
| No. of nonzero coefficients | $K \in \{4,8,16\}$ |
| Distribution of nonzero coefficient | Random Gaussian $\mathcal{CN}(0,1)$ |
| Received SNR for channel estimation | $SNR = 10\text{dB}\}$ |
| GMM noise distribution | $\alpha_1 = \alpha_2 = 0,\ \sigma_1^2 = 10^{(-SNR/10)}$ $\sigma_2^2 = T\sigma_1^2, T \in \{200,400,600\}$ |
| Step-size | $\mu = 0.01$ |
| Threshold of the (S)LMS-RL1 | $\delta_r = 0.05$ |

In the first example, average MSE curves of the proposed algorithm are depicted under different channel sparsity, i.e., $K \in \{4,8,16\}$ as shown in Figs. 1~3. Three figures show that MSE curves depend highly on regularization parameter $\lambda$. Under the simulation environment as listed in Tab. I, Fig. 1 shows that $\lambda = 4 \times 10^{-2}$ is feasible parameter for channel sparsity $K = 4$ while Figs. 2~3 demonstrate that $\lambda = 8 \times 10^{-2}$ is suggested parameter for $K \in \{8,16\}$. In practical system scenarios, channel sparsity ($K$) is often changed randomly. Hence, stability of channel estimation algorithm is the most important for selecting regularization parameter empirically. Considering the three representative cases $K \in \{4,8,16\}$ as shown in Figs. 1~3, $\lambda = 8 \times 10^{-3}$ is selected as for the SLMS-RL1 which can ensure stable convergence while without sacrificing significant MSE performance.

In the second example, average MSE curves of the proposed algorithm with regularization parameter $\lambda = 8 \times 10^{-3}$ are depicted under GMM impulsive-noise parameters, i.e., $T \in \{200,400,600\}$ as shown in Fig. 4. Under the certain circumstance, e.g., $SNR = 10$dB, GMM noise mixture parameter $\phi = 0.1$ as well as channel sparisty $K = 8$, one can find that proposed SLMS-RL1 is better than the state-of-the-art three algorithms under GMM noise with positive mixture parameters ($\phi$). In Fig. 4, MSE curves of LMS-types algorithms are decided by the different parameters ($T$). In other words, LMS-type algorithms are sensitive to $T$. In turn, SLMS-type algorithms are stable to different impulsive-noise parameters, i.e., $\in \{200,400,600\}$. The main reason of SLMS-type algorithms is utilized the sign function which is stable impulsive noise. Hence, the proposed algorithm is also stable for different GMM mixture parameters ($\phi$).

In the third example, average MSE curves of the proposed algorithm are depicted under different channel sparsity, i.e., $K \in \{2,4,8,16\}$ as shown in Fig. 5. Under certain circumstance, e.g., $\lambda = 8 \times 10^{-3}$, $SNR = 10\text{dB}$, GMM noise with impulsive-noise parameter $T = 400$ as well as mixture parameter $\phi = 0.1$, one can find that the proposed SLMS-RL1 is better than the state-of-the-art three algorithms under different channel sparsity $K$. In addition, one can find also that convergence speed of adaptive sparse algorithms (i,e, RL1-LMS and RL1-LAE) depends on $K$ and steady-state MSE curves of corresponding algorithms are very close. For different channel sparsity, in other words, the adaptive sparse algorithms may differ from conventional compressive sensing based sparse channel estimation algorithms [13], [20]–[22] which depend highly on channel sparsity. Hence, the proposed algorithm is also stable for different channel sparsity.

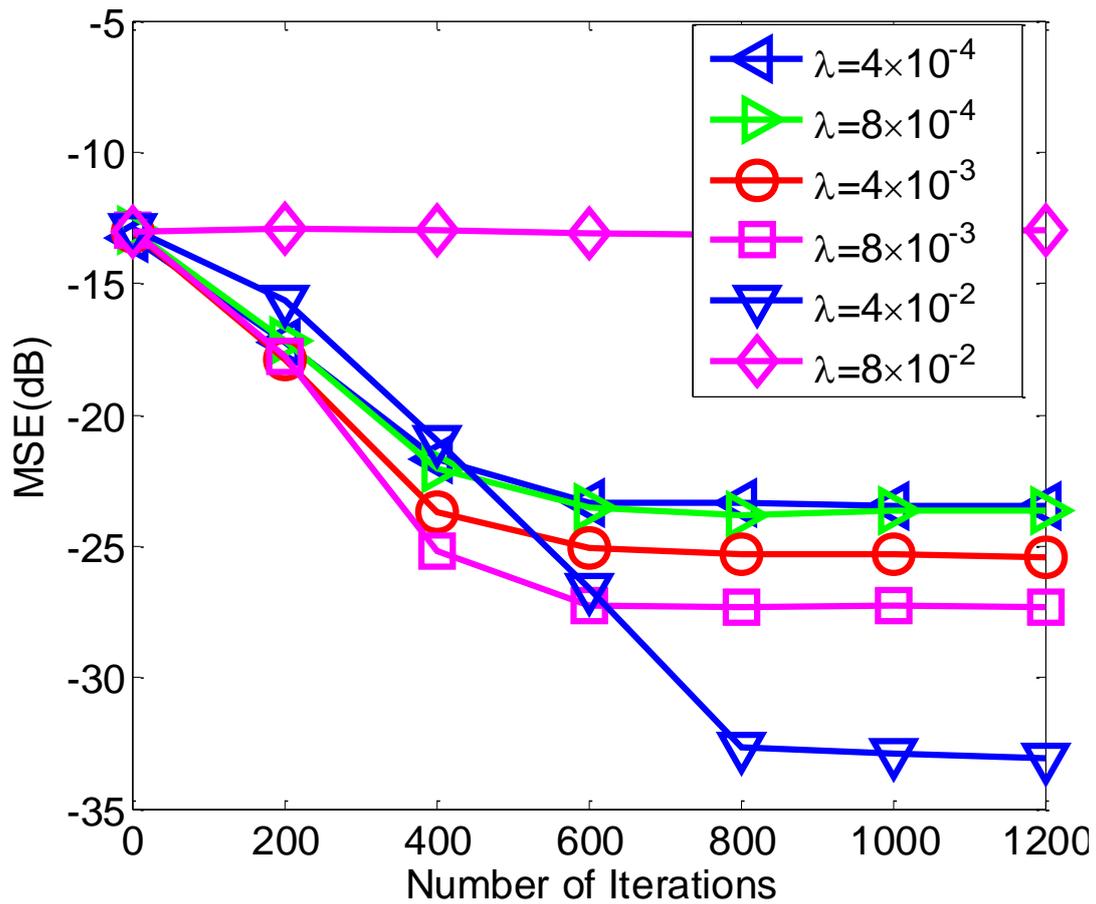

Fig. 1. Monte Carlo simulations averaging over 1000 runs for the channl sparsity $K = 4$, GMM with the mixture parameter $\phi = 0.1$ and the impulsive-noise strength $T = 400$ with respect to reguarlization parameter λ.

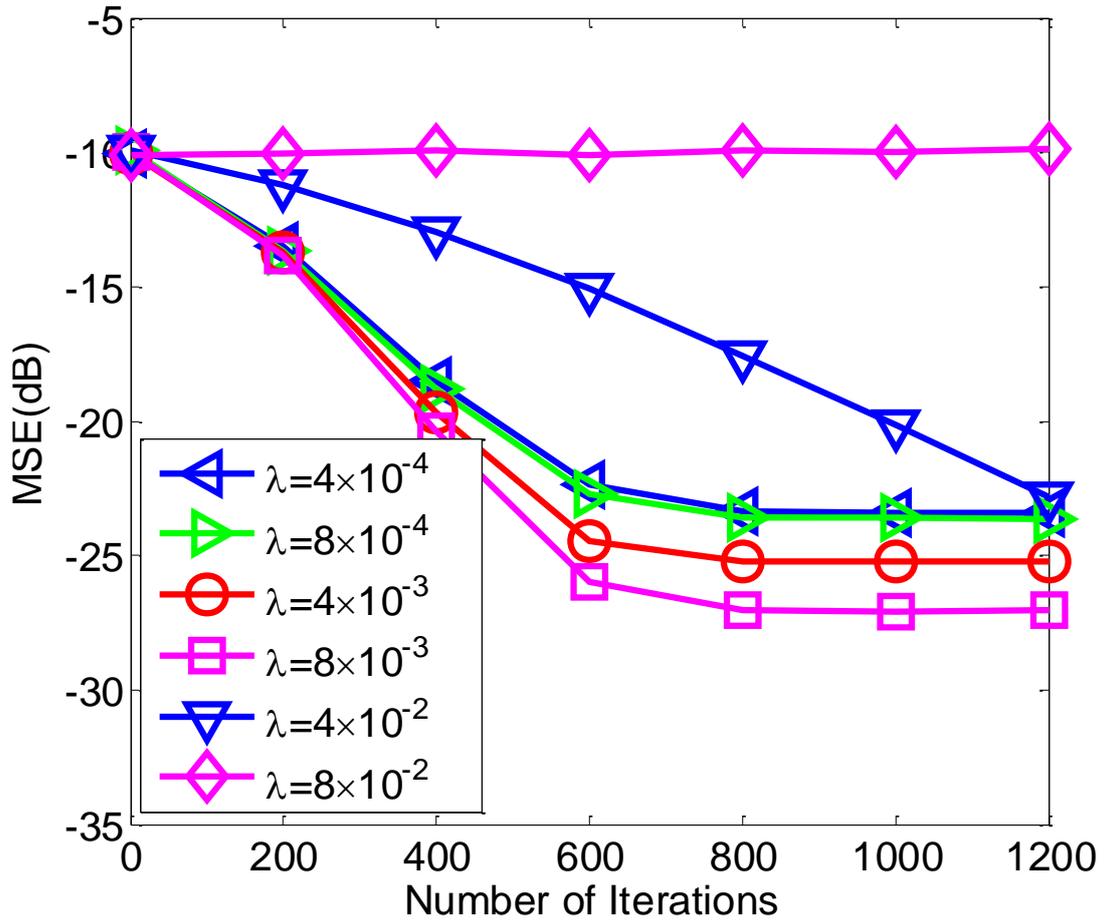

Fig. 2. Monte Carlo simulations averaging over 1000 runs for the channl sparsity $K = 8$, GMM with the mixture parameter $\phi = 0.1$ and the impulsive-noise strength $T = 400$ with respect to reguarlization parameter λ.

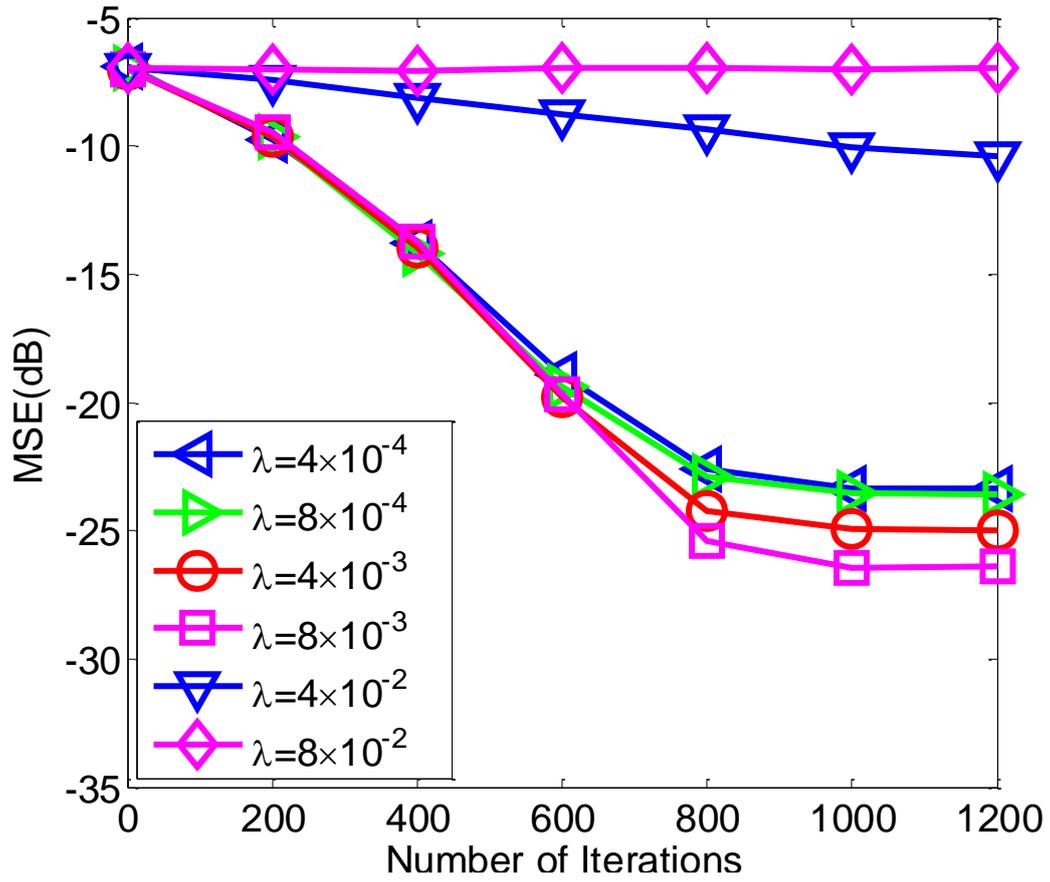

Fig. 3. Monte Carlo simulations averaging over 1000 runs for the channl sparsity $K = 16$, GMM with the mixture parameter $\phi = 0.1$ and the impulsive-noise strength $T = 400$ with respect to reguarlization parameter λ.

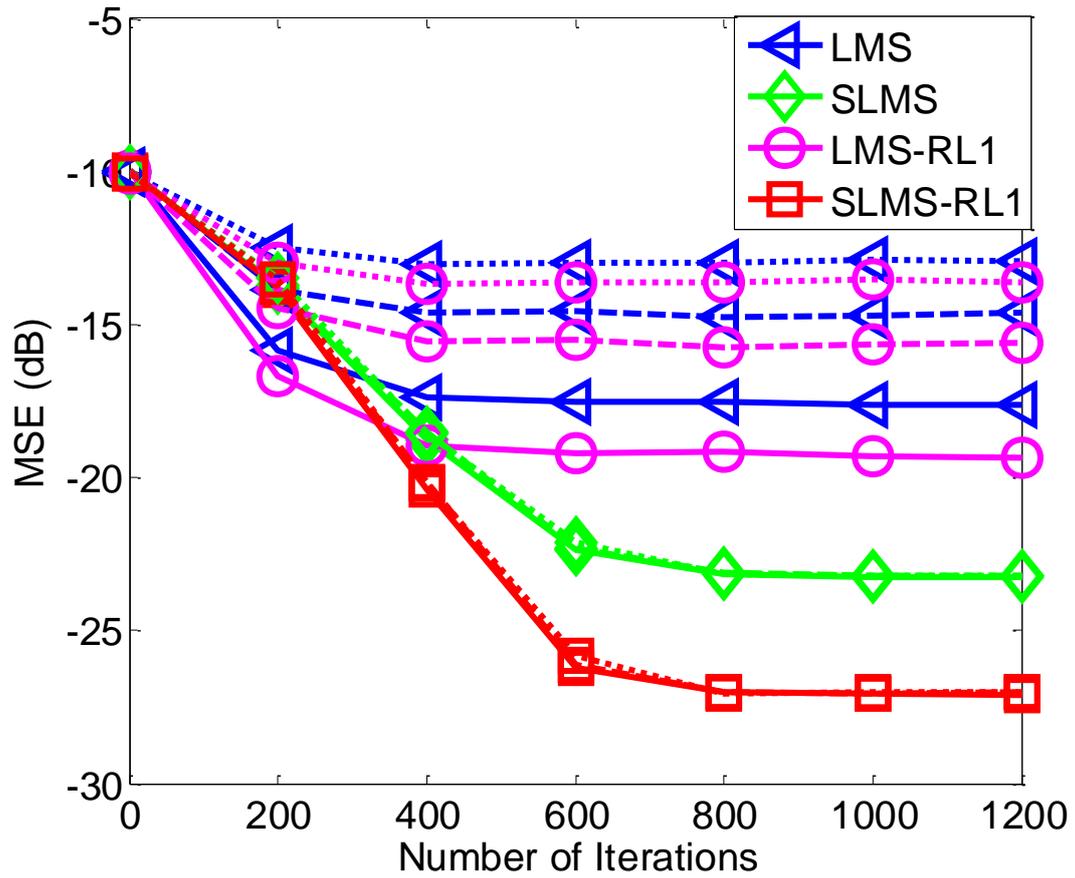

Fig. 4. Monte Carlo simulations averaging over 1000 runs for with the mixture parameter $\phi = 0.1$, the regularization paramter $\lambda = 8 \times 10^{-3}$, the channl sparsity $K = 8$, $SNR = 10$dB with respect to $T \in \{200, 400, 600\}$. Case 1 ($T = 200$): solid curves. Case 2 ($T = 400$): dashed curves. Case 3 ($T = 600$): dotted curves.

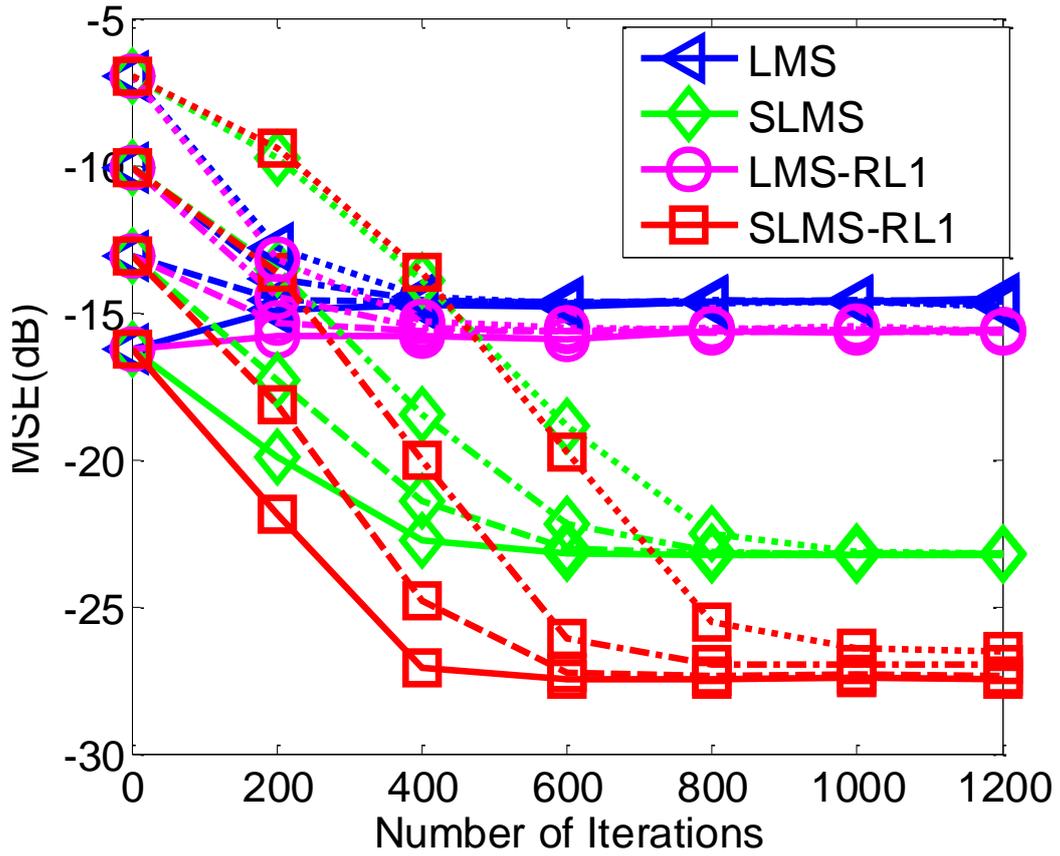

Fig. 5. Monte Carlo simulations averaging over 1000 runs for with the mixture parameter $\phi = 0.1$, the regularization paramter $\lambda = 8 \times 10^{-3}$, the channl sparsity $K \in \{2,4,8,16\}$, $SNR = 10$dB in $T = 400$. Case 1 ($K = 2$): solid curves. Case 2 ($K = 4$): dashed curves. Case 3 ($K = 8$): dashed-dotted curves. . Case 4 ($K = 16$): dotted curves.

## 4. Conclusions

Monte Carlo based REPA selection based SLMS-RL1 algorithm was proposed to estimate sparse channels under GMM environments. Considered three kinds channel sparsity, without loss of generality, $\lambda = 8 \times 10^{-3}$ was selected for SLMS-RL1 algorithm to exploit channel sparisty dependably as to ensure convergence stably. Simulation results demonstrated that the proposed algorithm obtained at least 5dB performance gain than the conventional LMS-RL1 algorithm with respect to different GMM noise strength ($T$) and different channel sparsity ($K$), respectively.

This paper only considered a simple scenario of applying the proposed algorithm to estimation sparse channels. The unknown channel dimension is often up to a few of hundreds. It is very difficult to apply the proposed SLMS-RL1 algorithm directly in higher-order dimensional (e.g., thousands or even higher) system identification. Based on the existing stable algorithms, in future work, we plan to develop kernel adaptive filtering [23], [24] based SLMS-RL1 algorithms which can deal with high-dimensional signal processing under non-Gaussian noise environments.

## 5. Competing interests

The authors declare that they have no competing interests.

## 6. Acknowledgements

This work was supported in part by Japan Society for the Promotion of Science (JSPS) research grants (No. 26889050 and No. 15K06072) as well as the National

Natural Science Foundation of China grants (No. 61401069, No. 61261048, No. 61201273).